\documentclass[journal=jacsat,manuscript=article]{achemso}

\usepackage[version=3]{mhchem} % Formula subscripts using \ce{}
\usepackage[utf8]{inputenc}
\usepackage{xcolor}
\usepackage{longtable}
\usepackage{multirow}
\usepackage{hyperref}
\usepackage{textgreek}
\usepackage{gensymb}
\usepackage[T1]{fontenc}
\usepackage{newtxtext,newtxmath}
\usepackage[mode=text]{siunitx}

\author{Hicham Moutaabbid}
\author{Dario Taverna}
\affiliation{IMPMC, Sorbonne Universit\'e, CNRS, MNHN, 4, place Jussieu, 75005 Paris, France}
\author{Denis Pelloquin}
\affiliation{Universit\'e de Normandie, UNICAEN, ENSICAEN, Laboratoire CRISMAT UMR CNRS 6508, 14050 Caen, France}
\author{Lorenzo Paulatto}
\affiliation{IMPMC, Sorbonne Universit\'e, CNRS, MNHN, 4, place Jussieu, 75005 Paris, France}
\author{Alexandre Gloter}
\author{Sophie Gu\'eron}
\author{Alik Kasumov}
\affiliation{Laboratoire de Physique des Solides, Universit\'e Paris-Saclay, 91400 Orsay, France}
\author{Andrea Gauzzi}
\affiliation{IMPMC, Sorbonne Universit\'e, CNRS, MNHN, 4, place Jussieu, 75005 Paris, France}
\email{andrea.gauzzi@sorbonne-universite.fr}

\title[]{A metallic CrS$_2$ phase bridging the gap between two- and three-dimensional dichalcogenides}

\abbreviations{IR,NMR,UV}

\begin{document}

\begin{tocentry}

Some journals require a graphical entry for the Table of Contents.
This should be laid out ``print ready'' so that the sizing of the
text is correct.

Inside the \texttt{tocentry} environment, the font used is Helvetica
8\,pt, as required by \emph{Journal of the American Chemical
Society}.

The surrounding frame is 9\,cm by 3.5\,cm, which is the maximum
permitted for  \emph{Journal of the American Chemical Society}
graphical table of content entries. The box will not resize if the
content is too big: instead it will overflow the edge of the box.

This box and the associated title will always be printed on a
separate page at the end of the document.

\end{tocentry}

\begin{abstract}
We report on the high-pressure synthesis of a CrS$_2$ phase in the form of single-crystalline nanorods. A structural refinement of Precession Electron Diffraction Tomography data confirms the nominal CrS$_2$ composition and unveils a ladder-type structure formed by portions of 1T-type CrS$_2$ layers characteristic of two-dimensional (2D) dichalcogenides connected by chains of edge-sharing CrS$_6$ octahedra characteristic of 3D dichalcogenides with marcasite structure. \textit{Ab initio} density functional theory calculations of the relaxed structure confirm the stability of this structure and indicate a strong overlap of the 3$d$ states of Cr with the 3$p$ states of S, thus suggesting strong covalent Cr–S bonds and metallic behavior. Electrical resistivity, $\varrho$, measurements on single nanorods confirm this behavior and yield $\varrho \sim$ 2-20 m$\Omega$ cm at 4 K. The proposed ladder-like structure of CrS$_2$ forms open channels along the chain direction, which may be suitable for ionic conduction.
\end{abstract}

\section{Introduction}
Owing to their tunable electronic, transport and optical properties, layered transition metal disulfides $M$S$_2$ ($M$ transition metal) have been the object of intense studies for decades \cite{wie80,wan12}. The origin of this tunability resides in the possibility of doping the $M$S$_2$ layers, which is achieved effectively by chemical intercalation or gating \cite{wu23} owing to the weak van der Waals cohesive forces between adjacent layers. As compared to their counterpart layered transition metal oxides, the amphoteric character of sulfur and the compressibility of the sulfur anion offer important advantages, such as mechanical flexibility and an improved structural stability during redox reactions and ionic transport \cite{gao13}. As a result, layered $M$S$_2$ compounds are promising for a variety of applications including energy storage, e.g. batteries \cite{gao13}, photonics \cite{mak16} and catalysis \cite{fu17}.

As previously discussed \cite{fan97} and explained by \textit{ab initio} calculations \cite{ray97,ray97b}, the progressive decrease of the energy of the $d$-band with filling in layered $M$S$_2$ structures like the 1T-, 2H- or 3R-types \cite{wie80} tend to reduce the $M^{4+}$ ions. As a result, these layered structures are stable only at low filling in compounds of the IV and V groups, while the three-dimensional (3D) pyrite or marcasite structures become stable at higher fillings, e.g. starting from group VII. In compounds of group V, layered structures are formed at ambient pressure only by the $4d$ and $5d$ ions Nb and Ta, while the layered 1T-VS$_2$ phase is metastable \cite{kat79,pod02} and obtained either by Li de-intercalation of LiVS$_2$ \cite{mur77} or under high pressure \cite{gau14,mou16}. By further increasing the filling, e.g. in the compounds of group VI, layered $M$S$_2$ structuress are again formed only by the $4d$ and $5d$ ions Mo and W while, to the best of our knowledge, a thermodynamically stable CrS$_2$ phase has not been reported yet either as 2D or 3D compound. Previously reported Cr sulfides typically host Cr$^{3+}$ or Cr$^{2+}$ ions, as in Cr$_2$S$_3$ and CrS \cite{jel57}, or intermediate valence Cr ions, as in Cr$_6$S$_7$ \cite{yuz83} or Cr$_5$S$_8$ \cite{sle69}. CrS$_2$ layers are found in ternary systems where the nominal valence of Cr is Cr$^{3+}$, such as the misfit-layer compound (LaS)$_{1+\alpha}$CrS$_2$ \cite{wie96,pan14}, $A$CrS$_2$, with $A$=K, Na, Cu, Ag or Au \cite{bon68,eng73,fuk99,nav18}, BaCr$_4$S$_7$ and Ba$_2$Cr$_5$S$_{10}$ \cite{fuk07}. Similar considerations apply to the layered compound 1T-CrSe$_2$ \cite{tan18}, where the valence of Cr is reduced owing to the weak ionic properties of selenides \cite{kob15}. A recent report on the synthesis of flakes of CrS$_2$ thin films \cite{pal17} suggests the possibility of synthesizing a layered CrS$_2$ phase under non-equilibrium conditions. 

Here, we attempt to stabilize the Cr$^{4+}$ ion in a bulk CrS$_2$ phase by using high-pressure, following a previous successful attempt to stabilize V$^{4+}$ in the 1T-VS$_2$ phase \cite{gau14,mou16}. Indeed, pressure is known to favor high-valence states of transition metals. Our motivation is that a bulk CrS$_2$ phase may contribute to a better understanding of the stability of transition metal dichalcogenides, a prerequisite to master redox reactions in new materials, a relevant issue for a variety of fields, such as energy-storage applications and catalysis. Our attempts have led to the synthesis of a CrS$_2$ phase that exhibits a ladder-type structure formed by portions of layered 2D 1T-type and 3D marcasite-type structures typical of $M$S$_2$ compounds at low- and high-filling levels, respectively.

\section*{Experimental and calculation methods}

\subsection*{High-pressure synthesis}
We carried a series of synthesis in a Paris-Edinburgh press under pressures of 4-5 GPa at variable temperatures in the 400-900 $^{\circ}$C range. A typical synthesis lasted one hour, followed by a quenching of the sample down to room temperature to prevent the formation of competing phases and by a slow release of pressure. The samples consist of a cylindrical capsule made of a 25 $\mu$m thick Pt foil or an hexagonal BN sleeve filled by the precursor powders. The capsule is inserted into a hollow cylinder of graphite serving as furnace. We used two different precursors. 1) A mixture of Cr$_2$S$_3$ and S powders (Alfa Aesar, 99.9995\%) in the nominal 1:1 ratio. The goal was to follow a previous study \cite{mou16} that enabled the high-pressure synthesis of nearly stoichiometric 1T-VS$_2$ by de-intercalating the V ions in the van der Waals gap between VS$_2$ layers in a V$_5$S$_8$ precursor, thus stabilising V$^{4+}$ ions. We expected that the same process may occur in Cr$_2$S$_3$ as well, for this structure also contains Cr ions intercalated between 1T-type CrS$_2$ layers. 2) A 1:2.2 mixture of metallic Cr powders (Sigma-Aldrich, 99.9\%, 325 mesh) and S powders. Our motivation was that an hypothetical 1T CrS$_2$ phase would be favored under pressure considering that the density of this phase is higher than the density of the precursor. The excess of S with respect to the 1:2 nominal composition was meant to form a sulfur flux, which is expected to favor the growth of single crystals. 

\subsection*{Structural characterization}
The morphology and chemical composition of the as-prepared samples were investigated using a Zeiss Ultra 55 Scanning Electron Microscope equipped with a Field Electron Gun (FEG-SEM) and a Bruker 125 eV AXS energy dispersive X-ray spectroscopy (EDS) analyzer. The crystal structure was determined by means of High-Resolution Transmission Electron Microscopy (HREM) by refining Precession Electron Diffraction data recorded in Tomography mode (PEDT) combined to structural projection images obtained using HREM and High-Angle Anular Dark Field Scanning Transmission Electron Microscopy (HAADF)/STEM. The STEM images were recorded with a corrected probe JEOL ARM-200 operating at 200 kV and equipped with a ADF (Anular Dark Field) detector. Simulated ADF images have been calculated using the JEMS software considering the convolution of the STEM probe with the intensity of the object (square of the projected potential multiplied by the electron-matter interaction constant of the structure).

In order to determine the structure of the samples, we have carried out a diffraction study on individual nanocrystals using a JEOL 2010Cx microscope operating at 200 kV (point resolution 2.7 \AA) equipped with a DIGISTAR module for the precession. The samples were prepared by grinding the bulk material, dispersing the resulting powder in ethanol using ultrasonic bath and depositing it onto a copper grid with a holey carbon film. The data were collected with a Gatan ORIUS D200 camera. Manual electron diffraction (ED) tomography was performed using a tilt-rotating sample holder and applying an angular tilt step of 1$^{\circ}$. The data set was collected with a 800 mm camera length and a +53$^{\circ}$/–48$^{\circ}$ tilt range. A precession angle of 1$^{\circ}$ was applied to integrate the whole intensity of the reciprocal space rods. No alteration of the samples induced by the electron beam was detected. The data processing was performed using the PETS software developed by Palatinus \textit{et al.} \cite{pal17}. Data reduction, cell refinement, space group determination, scaling corrections were performed using the Jana2006 package \cite{pet14}. The structure was solved using the charge flipping method (Superflip program) implemented in Jana2006. The final structure and refinements were carried out by Fourier difference maps using Jana2006. 

\subsection*{EELS study}
High resolution scanning transmission electron microscopy (STEM) and electron energy loss spectroscopy (EELS) analysis were performed using a Cs aberration-corrected STEM, the NION UltraSTEM200 operated at 100 kV and coupled with a high-sensitivity EEL spectrometer. The samples were prepared as above.

\subsection*{Transport Measurements}
We measured the electrical resistance of single CrS$_2$ nanorods using a two-terminal method; the usual four-terminal method would be not feasible owing to the smallness of the nanorods. The single nanorods to be measured were selected by dissolving a tiny amount of powder in chloroform and dispersing the powder using ultrasounds. A drop of solution was deposited on an oxidized Si substrate and left to dry. As seen in Figure \ref{fig:transport}, each terminal consists of rather macroscopic Nb electrode in series with a 100 nm-thick W wire in direct contact with the selected CrS$_2$ nanorod. Following the preparation route described in detail elsewhere \cite{kas05}, the Nb electrodes were prepared by e-beam lithography on previously sputtered Nb films, while the W wires were prepared by Focused Ion Beam. The method exploits the superconducting properties of Nb and W, which enables us to measure only the resistance of the CrS$_2$ nanorod and of a possible rod/electrode contact resistance, and thus to estimate the resistivity of the nanorod at low temperatures, where the resistance of the electrodes is zero. We successfully deposited a pair of Nb/W electrodes on ten segments of single CrS$_2$ nanorods. Typical length of the segments was 1-2 $\mu$m. The width of the nanorods varies in the 20-300 nm range, depending on the number of nanorods forming the wire bundle. As reference sample, using FIB, we fabricated a W nanowire connecting the Nb electrodes, in the place of the CrS$_2$ sample. This reference sample was used to test the zero-resistance of the two-terminal device.

\subsection*{Ab initio calculations}
We calculated the relaxed structure in the P1 group space starting with the experimental C/2$m$ structure (see Results section) as input. We used the Perdew-Burke-Ernzerhof (PBE) functional \cite{per96} of the Generalized Gradient Approximation (GGA) of Density Functional Theory (DFT) as implemented in the Quantum Espresso code \cite{gia09,gia17,gia20} and the Garrity-Bennett-Rabe-Vanderbilt (GBRV) ultra-soft pseudopotentials \cite{gar14} with a 40 Ry kinetic-energy cutoff. We have chosen GGA for this approximation tends to yield more accurate total energies and thus better discriminates between competing structures, a crucial aspect of the present study, as compared to the Local Spin Density approximation. In order to take into account the van der Waals interaction, which plays a fundamental role in stabilizing the crystal structure of layered dichalcogenides, we have adopted the semi-empirical DFT+D2 correction proposed by Grimme \cite{gri06} that introduces an additional damped ion-ion $\sim R^{-6}$ force and produces no direct effect on the electronic structure. The electronic band structure was calculated in a 4 $\times$ 4 $\times$ 2 symmetry-reduced $k$-points of the first Brillouin Zone and a 0.01 Ry Marzari-Vanderbilt-De Vita-Payne cold smearing \cite{mar99} to integrate up to the Fermi surface.

\section{Results and discussion}
\subsection*{Morphology and composition of the nanocrystals}
\begin{figure}[htb!]
	\centering
    \includegraphics[width=0.75\textwidth]{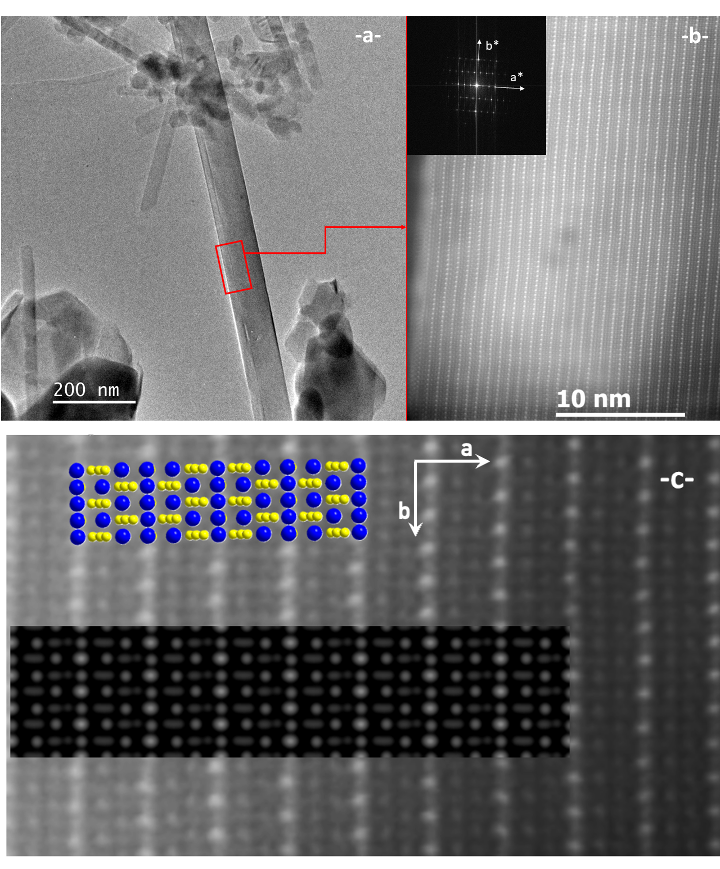}
	\caption{a) TEM image of a bundle of CrS$_2$ nanorods mixed with nanocrystals of different morphology. b) High-resolution HAADF image along the [001] direction recorded from an individual nanorod and corresponding Fast Fourier Transform (FFT) pattern. c) An enlarged portion of the HAADF image is compared with a simulated image calculated from the structural model obtained from the PEDT data analysis, as described in the text. The schematic structure projected along the [001] direction is also shown. Cr and S atoms are represented in blue and yellow, respectively.}
	\centering
	\label{fig:bundle}
\end{figure}

Using the first precursor, i.e. a stoichiometric mixture of Cr$_2$S$_3$ and S powders, and in the whole temperature and pressure ranges investigated, as described in the preceding section, we always obtained polycrystalline samples containing CrS as main phase. Using the second precursor, i.e. a mixture of metallic Cr and S, we obtained polyphasic samples with heterogenous morphology. We focused our attention on the samples synthesized at 5 Pa and 600 $^{\circ}$C, where SEM and TEM analysis unveiled the presence of micro- and submicro-metric grains including nanorods of variable lengths up to 20 $\mu$m and of variable width in the 30-200 nm range (see Figure \ref{fig:bundle}a). Longer synthesis times do not affect the size of the nanorods and the same results were obtained using hexagonal BN or Pt cells. The morphology of the nanorods is similar to that of CrO$_2$ nanorods \cite{dho10} and completely different from the typical morphology found in previously reported chromium sulfides, such as petal-like crystals of Cr$_2$S$_3$ and hexagonal platelets of CrS$_x$ \cite{nar87}. We determined the chemical composition of three representative nanorods by means of EDX spectroscopy. For each crystal, we have taken a EDX spectrum on ten distinct spots and consistently found a Cr:S ratio of 1:2 within the experimental error by using FeS$_2$ and metallic Cr single crystals as standards. No trace of impurity element was detected in any of the nanorods within the limit of resolution of EDX spectroscopy.

\subsection*{Experimental determination of the CrS$_2$ structure}
Given the heterogeneous morphology and the sub-micrometric size of the crystals forming the polycrystalline samples, we employed Precession Electron Diffraction (PED) \cite{pal17} as the method of choice to determine the crystal structure of individual CrS$_2$ nanorods. According to this method, we extracted the intensity, $I$, of the diffraction peaks from the whole PED patterns and compiled the data together in order to build the experimental 3D reciprocal lattice. We first attempted to refine the data assuming a triclinic P$\bar 3 m$1 structure similar to the 1T structure of CrSe$_2$ \cite{kob14}, but it was evident that the data are not compatible with this structural model. We then explored alternative structural models and finally found a good match of the data with a monoclinic cell.

\begin{figure}[htb!]
	\centering
    \includegraphics[width=0.9\textwidth]{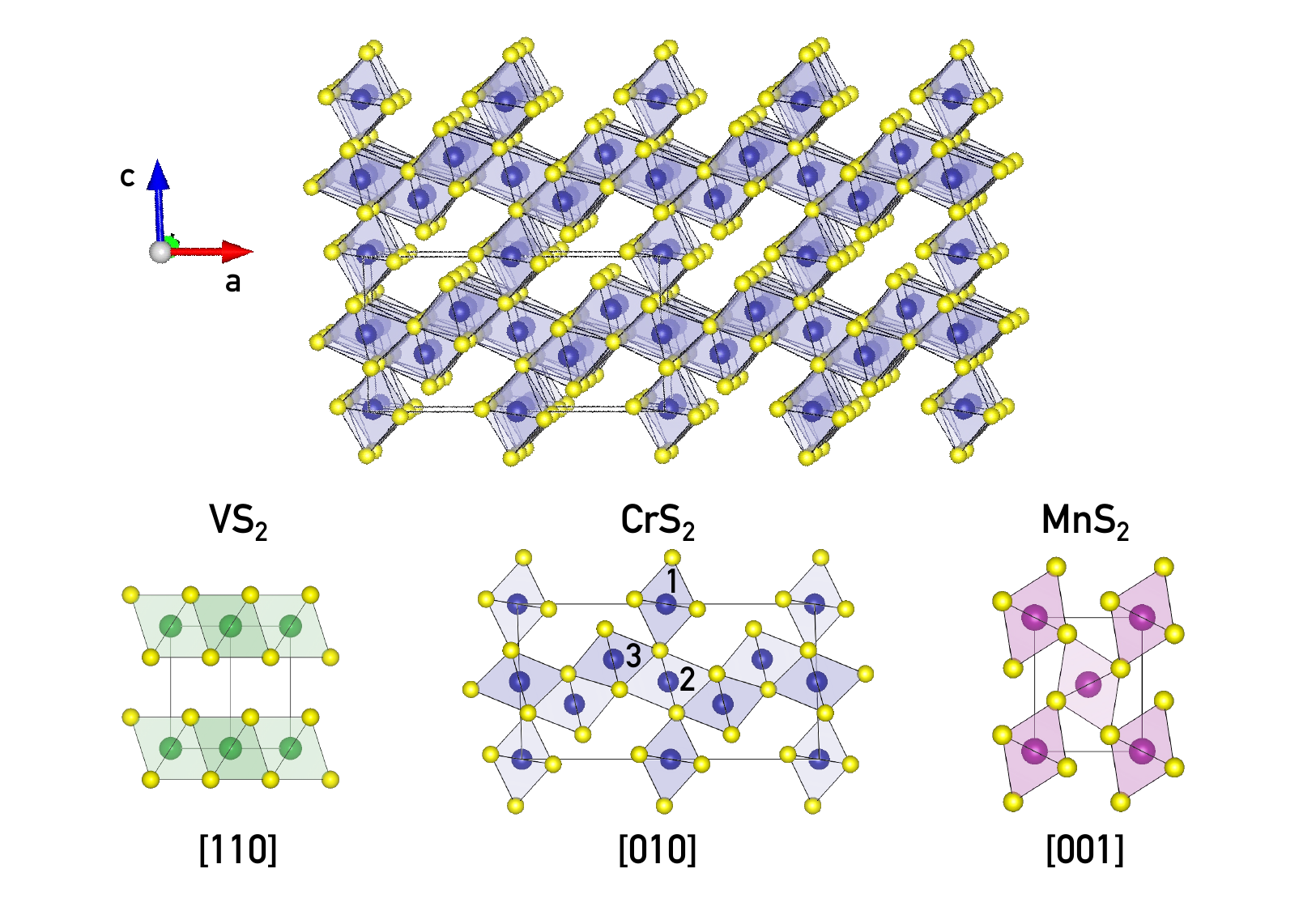}
	\caption{Top: monoclinic C2/$m$ structure of CrS$_2$ obtained by refining the precession electron diffraction data and supported by \textit{ab initio} calculations, as described in the text. Blue and yellow spheres represent Cr and S atoms, respectively. The black solid line indicates the unit cell. The structural parameters are reported in Table \ref{tab:structure}. Note a ladder structure where the steps are portions of 1T-type layers connected by chains of edge-sharing CrS$_6$ octahedra along the $b$ [010] direction. Bottom: for comparison, we show the trigonal P$\bar 3m$1 1T structure of VS$_2$ \cite{mur77}, formed by layers of edge-sharing octahedra, and the orthorhombic P$nnm$ marcasite structure of MnS$_2$ \cite{cha85}, formed by chains of edge-sharing octahedra along the $c$ direction. Both building blocks are present in the CrS$_2$ structure as well. The labels 1, 2, 3 indicate the three Cr sites of the CrS$_2$ structure in Table \ref{tab:structure}.}
	\centering
	\label{fig:structure}
\end{figure}

For the next step of structure determination, we retained 690 independent reflections of intensity satisfying the $I > 3\sigma$ criterion among the 1784 measured reflections measured. Using the JANA2006 software, we selected the C2/$m$ space group that gave the best redundancy of 2.749 and $R_{\rm int}$=21.98\%. After 2000 cycles of calculations, SUPERFLIP found a solution with a cumulative coverage of 100\%, a $d_{min}$ resolution value of 1 \AA\ and an overall agreement factor of 10.76\%. We consider this result to be satisfactory taking into account that dynamical diffraction effects, which are comparatively important in electron diffraction data, are not considered in the refinement. A refinement of the room temperature structure by means of electron diffraction confirms the C2/$m$ structure and yields the following lattice parameters: $a$ = 14.38(1) \AA, $b$=3.391(2) \AA, $c$ =7.514(8) \AA, $\beta$=91.68(7)° with reliability factors $R_1$= 26.79\% ($I > 3\sigma$) and $wR_2$= 34.71\%. We give the parameters of the refined structure including the atomic positions in Table \ref{tab:structure} and schematically show the structure in Figure \ref{fig:structure}. The validity of the refinement is supported by the excellent match between the simulated HAADF image of the refined structure and an experimental high-resolution HAADF image of a representative CrS$_2$ nanorod along the [001] direction (see Figure \ref{fig:bundle}).

\renewcommand{\arraystretch}{1.1}
\setlength{\tabcolsep}{10pt}

\begingroup

\begin{table}[htb!]
\caption{Experimental and calculated structures of CrS$_2$ in the monoclinic C2/$m$ symmetry. The former structure has been refined from the PEDT data, while the latter structure is obtained by approximating the structure calculated in the P1 symmetry in the monoclinic C2/$m$ pseudosymmetry (see text). Numbers in parentheses indicate statistical uncertainty of the experimental values.}
\label{tab:structure}
\begin{tabular}{|| c c | l l l | l l l ||}
\hline \hline
\multicolumn{2}{|| c |}{Lattice parameters} & \multicolumn{3}{c |}{Experimental} & \multicolumn{3}{c ||}{Calculated} \\
\hline
\multicolumn{2}{|| c |}{$a$ [\AA]} & \multicolumn{3}{c |}{14.38(1)} & \multicolumn{3}{c ||}{14.5653} \\
\multicolumn{2}{|| c |}{$b$ [\AA]} & \multicolumn{3}{c |}{3.391(2)} & \multicolumn{3}{c ||}{3.1506} \\
\multicolumn{2}{|| c |}{$c$ [\AA]} & \multicolumn{3}{c |}{7.514(8)} & \multicolumn{3}{c ||}{8.0724} \\
\multicolumn{2}{|| c |}{$\beta$ [\degree]} & \multicolumn{3}{c |}{91.68(7)} & \multicolumn{3}{c ||}{92.340} \\
\multicolumn{2}{|| c |}{$V$ [\AA]$^3$} & \multicolumn{3}{c |}{366.2(9)} & \multicolumn{3}{c ||}{370.129} \\
\hline \hline
Site & Wyckoff position & $x$ & $y$ & $z$ & $x$ & $y$ & $z$ \\
\hline
Cr1	& $2b$ & 0.5       &  0   & 0        & 0.5	   & 0  &  0 \\ 
Cr2	& $4i$ & 0.6807(6) &  0   & 0.360(1) & 0.6817  & 0  &  0.3555 \\
Cr3	& $2c$ & 0.5	   &  0.5 & 0.5      & 0.5     & 0.5 & 0.5 \\ 
S1	& $4i$ & 0.660(1)  &  0.5 & 0.556(2) & 0.6642  & 0.5 & 0.5498 \\ 
S2  & $4i$ & 0.528(1)  &  0	  & 0.292(2) & 0.5256  & 0   & 0.2982 \\ 
S3  & $4i$ & 0.604(1)  &  0.5 & 0.978(2) & 0.6046  & 0.5 & 0.9681 \\ 
S4  & $4i$ & 0.703(1)  &  0.5 & 0.156(2) & 0.7076  & 0.5 & 0.1599 \\ 
\hline \hline
\end{tabular}

\end{table}

\endgroup

\subsection*{\textit{Ab initio} calculated CrS$_2$ structure}
The calculations of the optimized structure further support the validity of the structural model proposed. As customary, we have first calculated the relaxed structure that minimizes the total energy in the lowest P1 symmetry. Using the standard PBE functional described in the Experimental Methods, the calculated structure was found to differ significantly from the experimental one, as indicated by the 20\% larger volume of the relaxed unit cell as compared to the experimental one. We attribute this discrepancy to the fact that the standard PBE functional used does not include the van der Waals forces typically required to stabilize layered dichalcogenides. By including these forces as described in the Experimental Methods, the calculated P1 structure was found to match the experimental C2/$m$ one, as discussed below.

A first indication of this match is that the calculated and experimental unit cell volumes agree within 1\%. Second, using the PSEUDO \cite{cap11} and STRAIN \cite{aro06} programs of the Bilbao Crystallographic Server, we found that the calculated P1 structure has a C2/$m$ pseudosymmetry, consistent with the C2/$m$ symmetry of the experimental structure described in the previous section, with a degree of lattice distortion $S=3.7$, defined as the square root of the average of the squared eigenvalues of the strain tensor, $\varepsilon_i$. The magnitude of this distortion falls within the typical range of reliability of GGA results. Finally, as seen in Table \ref{tab:structure}, the calculated $a$, $b$, $c$ and $\beta$ lattice parameters of the approximated C2/$m$ unit cell agree with those of the refined C2/$m$ unit cell within 1\%, 7\%, 7\%, and 0.7\%, respectively. This discrepancy corresponds to a further small distortion, $S=0.035$, also compatible with the limitations of the GGA approximation that tends to expand and soften bonds \cite{pro95}. Interestingly, the refined atomic positions in relative lattice units agree very well with those of the refined structure within 2\% or better for all atomic sites. Therefore, the resulting differences in the atomic positions mainly arise from the aforementioned difference of unit cell parameters. On average, the atomic positions differ by $d_{\rm av}=0.1715$ \AA; the Cr2 site exhibits the largest difference of 0.3866 \AA, which is attributed to the lower symmetry of this site that leaves more freedom to its displacement in the calculation of the relaxed structure. With this \textit{caveat}, it is evident that the calculated structure matches well the refined one, thus supporting the structural model of Figure \ref{fig:structure}.

\subsection*{Discussion on the CrS$_2$ structure and on the valence state of Cr}

The proposed CrS$_2$ structure consists of a ladder-like structure where the steps are portions of 1T-type CrS$_2$ layers connected by chains of edge-sharing CrS$_6$ octahedra along the $b$-direction of the monoclinic cell. These chains are the building block of the 3D marcasite structure characteristic of dichalcogenides at higher filling of the $d$-band, e.g. MnS$_2$ and FeS$_2$. The structure forms open channels along the chain, which may be interesting for applications that require ionic conduction. Similar ladder-type structures containing shorter portions of CrS$_2$ layers have previously been reported in Er$_4$CrS$_7$ and Er$_6$Cr$_2$S$_{11}$ \cite{vaq09}. In conclusion, the crystal structure of the CrS$_2$ nanorods shares with both 1T- and marcasite-type dichalcogenides the same building blocks made of edge-sharing CrS$_6$ octahedra, as highlighted in Figure \ref{fig:structure}. The observation of an intermediate structure bridging the 2D and 3D structures found in transition metal sulfides is consistent with the intermediate filling level of the $d$-band of the Cr ion in CrS$_2$, which is too high to stabilize the 2D 1T structure and too low to stabilize the 3D marcasite structure. 

The stoichiometric CrS$_2$ composition of the nanorods, as determined by the previous EDX analysis and structural refinement, corresponds to a formal Cr$^{4+}$ valence state of the Cr ions. This comparatively high valence state for chromium sulfides is indeed consistent with the observation that the Cr-S bond lengths of CrS$_2$ are shorter than those of representative ambient pressure chromium sulfides that formally host Cr$^{3+}$ ions, such as Cr$_2$S$_3$ \cite{jel57}, CuCrS$_2$ and AgCrS$_2$ \cite{bon68}. Namely, the bond lengths of the above phases are 2.42, 2.39 and 2.36 \AA, respectively, while those of the three Cr1, Cr2 and Cr3 sites of the CrS$_2$ structure (see Figure \ref{fig:structure}) fall in the 2.26-2.38 \AA\ range (see Table \ref{tab:BVS}). Longer Cr-S bond lengths are also found in other high-pressure phases, such as Ba$_3$CrS$_5$ and Ba$_3$Cr$_2$S$_6$ \cite{fuk03}, with bond lengths in the 2.47-2.49 and 2.44-2.49 \AA\ range and formal Cr valences 4+ and 3+, respectively, or BaCr$_4$S$_7$ and Ba$_2$Cr$_5$S$_{10}$ \cite{fuk07}, with bond lengths in the 2.30-2.45 and 2.33-2.47 \AA\ range, respectively, and formal Cr valence 3+.

The above discussion indicates an average valence state significantly larger than 3+. We give a quantitative estimate by means of a bond valence sum (BVS) analysis according to the expression \cite{bro85}:

\begin{equation}
    v=\sum_{i=1}^6 \exp{\frac{R_0-R_i}{B}}
\end{equation}

where $v$ is the valence state of the Cr ion in a given site, i.e. Cr1, Cr2 or Cr3, the sum extends over the six Cr-S bonds of the CrS$_6$ octahedron, $R_0$ is an appropriate constant characteristic of Cr in a given valence state and in a given coordination number and taken from standard compounds, and $B$ is a constant, generally taken as $B=0.37$ \cite{bro85}. Using the value $R_0=2.162$ \AA\ reported by Brown \cite{bro17} and the experimental Cr-S bond length distances of Table \ref{tab:BVS}, we obtain valence states varying from 3.68+ for the Cr3 site to 4.72+ for the Cr1 site (see Table \ref{tab:BVS}). Using the calculated bond length distances, we find very similar BVS values for the Cr2 and Cr3 sites but a 0.55 smaller value for the Cr1 site. In spite of this discrepancy, which may be explained by the semi-quantitative character of the BVS method and by the lack of well-established $R_0$ values for the Cr-S bond \cite{bro17}, the present BVS analysis consistently indicates an average 4+ or larger valence state of Cr in CrS$_2$.
%The higher valence in Cr(1) site may be the connection to the CrS$_2$ sheet by the corner and not the face or the edge. Because of the strong Coulomb repulsion between Cr ions, ions Cr$^{4+}$ do not easily reside in face-sharing octahedra (see BaCr$_4$S$_7$).

\setlength{\tabcolsep}{7pt}
\renewcommand{\arraystretch}{1.2}
\begin{table} [htpb!]
\caption{Selected bond lengths in \AA, bond valence sum (BVS) values and bond angle variance, $\langle \Delta \vartheta^2 \rangle$, in deg$^2$ for the experimental and calculated structures of the previous table. BVS values are computed using $R_0$= 2.162 \r{A} and $B$ = 0.37 \AA\ \cite{bro17}.}
\centering
\begin{tabular}{ || c  c  c | c  c  c | c  c  c   || }
\hline \hline
\multicolumn{3}{|| c |}{} & \multicolumn{3}{c|}{Experimental} & \multicolumn{3}{c ||}{Calculated} \\
\hline \hline
Site & Bond & Mult. & Bond length & BVS & $\langle \Delta \vartheta^2 \rangle $ & Bond length & BVS & $\langle \Delta \vartheta^2 \rangle $ \\
 \hline
\multirow{2}{*}{Cr1} & Cr1-S2 & $\times$2 & 2.220(16) & \multirow{2}{*}{4.72} & \multirow{2}{*}{19.10} & 2.30239 & \multirow{2}{*}{4.17} & \multirow{2}{*}{12.44} \\
& Cr1-S3 & $\times$4 & 2.267(10) &  &  & 2.29378 &  & \\
\hline
\multirow{4}{*}{Cr2} & Cr2-S1 & $\times$1 & 2.356(16) & \multirow{4}{*}{4.23} & \multirow{4}{*}{11.77} & 2.76446 & \multirow{4}{*}{4.21} & \multirow{4}{*}{74.34} \\
& Cr2-S1 & $\times$2 & 2.271(13) &  &  & 2.26339 &  &  \\
& Cr2-S2 & $\times$1 & 2.233(16) &  &  & 2.09404 &  &  \\
& Cr2-S4 & $\times$2 & 2.314(13) &  &  & 2.32490 &  &  \\
\hline
\multirow{2}{*}{Cr3} & Cr3-S1 & $\times$2 & 2.330(14) & \multirow{2}{*}{3.68}& \multirow{2}{*}{13.96} & 2.33624 & \multirow{2}{*}{3.56}& \multirow{2}{*}{21.94} \\
& Cr3-S2 & $\times$4 & 2.350(11) &  &  & 2.36541 &  &  \\
                     
\hline \hline
\end{tabular}
\label{tab:BVS}
\end{table}

\begin{figure}[htb!]
	\centering
    \includegraphics[width=0.7\textwidth]{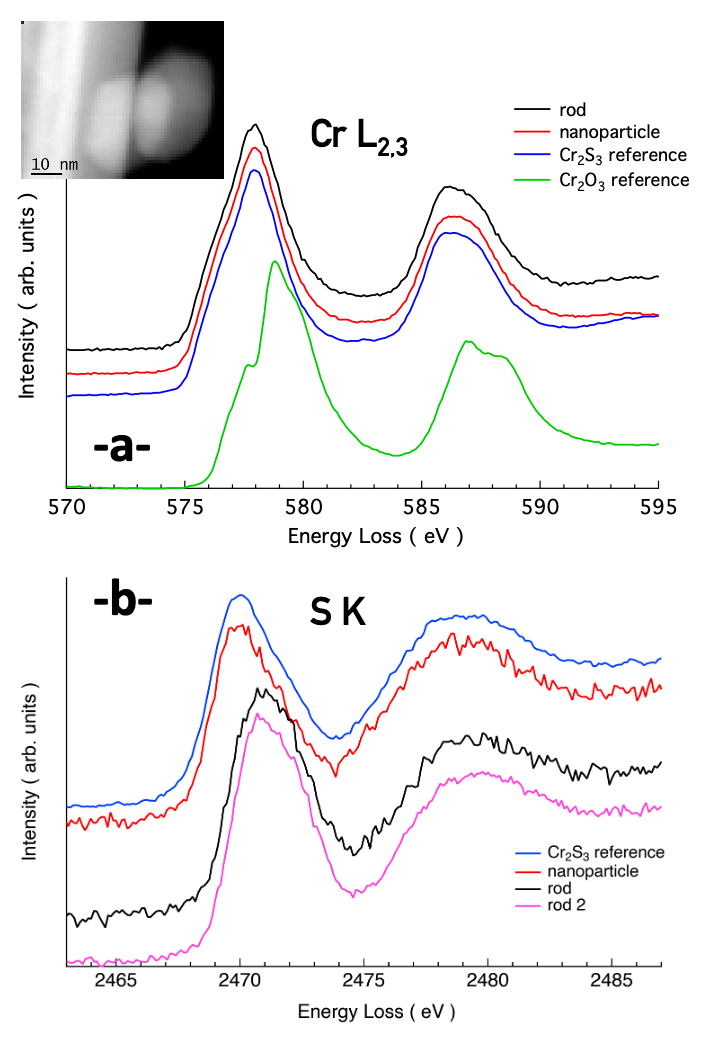}
	\caption{a) EELS Cr-L$_{2,3}$ absorption edge spectra of a representative CrS$_2$ nanorod and of a faceted Cr$_2$S$_3$ nanoparticle decorating the rod. The two objects are clearly recognized in the HAADF image in the inset. The spectra are compared with reference spectra of Cr$_2$S$_3$ and CrO$_2$. b) The same as in a) for the S-K absorption edge spectra. The spectrum of a second CrS$_2$ nanorod (labeled rod 2) is also shown to check the reproducibility of the result.}
	\centering
	\label{fig:XAS}
\end{figure}

\subsection*{EELS Cr-L$_{2,3}$ and S-K spectra}

The EELS results below confirm that the valence state of Cr in the CrS$_2$ nanorods is significantly larger than Cr$^{3+}$ ions. In Figure \ref{fig:XAS} we show the STEM-EELS Cr-L$_{2,3}$ and S-K absorption edge spectra of two representative nanorods and of a spurious nanoparticle attached to one nanorod. We compare these spectra with those of Cr$_2$S$_3$ and CrO$_2$ used as reference compounds for the Cr$^{3+}$ valence state with different ligands. Looking at the Cr-L$_{2,3}$ edges, we first note that the spectrum of CrO$_2$ differs for a better defined multiplet structure and for a 0.9 eV shift of the edge band toward higher energies, in agreement with a previous report by Daulton and Little \cite{dau06}. This shift is explained by the more ionic character of oxides as compared to sulfides. Second, the spectra of CrS$_2$ and of the impurity are very similar to Cr$_2$S$_3$. Specifically, the position of the absorption edge is unchanged within the experimental resolution. At first sight, this may indicate no change of valence state in CrS$_2$ as compared to Cr$_2$S$_3$. In fact, Suzuki and Tomita \cite{suz97} pointed out that transition metal ions in different valence states do not necessarily show different L spectra, as in the case of CrO$_2$ and Cr$_2$O$_3$ that exhibit very similar Cr-L$_{2,3}$ spectra, although they host Cr different Cr$^{4+}$ and Cr$^{3+}$ valence states, respectively. Instead, in the O-K edge spectrum of CrO$_2$, Suzuki and Tomita found a new peak at energies well below the absorption edge of Cr$_2$O$_3$, consistent with the presence of a hole with predominant $p$ character. This observation suggests a charge transfer 3d$^3\underline{L}$ electronic configuration for CrO$_2$, consistent with a 4+ valence, while the electronic configuration of Cr in Cr$_2$O$_3$ is primary ionic (3d$^3$). This picture explains the weak dependence of the L spectra on the valence state of Cr in these two oxides.

In order to verify whether a similar ligand-hole configuration applies to our case, in Figure \ref{fig:XAS} we compare the S-K absorption spectra of the same three samples as above, i.e. the CrS$_2$ nanorod, the impurity nanoparticle and the Cr$_2$S$_3$ standard, plus an additional nanorod to test reproducibility. In agreement with previous studies on other binary transition metal sulfides, i.e. MnS, FeS and CoS\cite{sug81,pon94,far00}, we found an edge band around 2470 eV, assigned to the transition of S 1s electrons to both unoccupied hybridized S 3p$\sigma^{\ast}$ antibonding and empty Cr 3d($e_g$) states, and a broader band around 2480 eV assigned to the transitions to p-like states at higher energies \cite{sug81}. The spectra of the nanoparticle and of the Cr$_2$S$_3$ standard are identical, which indicates that the nanoparticle is a Cr$_2$S$_3$ impurity. Instead, in the spectra of the CrS$_2$ nanorods, the edge band is shifted towards higher energies and the relative intensity of this band increases as compared to the high-energy band. According to a systematic study of S K-edge XAS spectra of transition metal sulfides by Fleet \cite{fle05}, this shift is attributed to an increase of the average oxidation state of sulfur in CrS$_2$, consistent with a higher valence state of Cr caused by a stronger ligand-hole contribution as compared to Cr$_2$S$_3$.  

%According to a systematic study of S-K XAS spectra of transition metal sulfides by Fleet \cite{fle05}, this shift is attributed to an increasing ionic character of CrS$_2$ as compared to Cr$_2$S$_3$, consistent with a scenario of higher valence state of Cr.  

%Soldatov et al. \cite{sol04}. In this study, the authors found the same sharp low-energy band and broad high-energy band as in our case and noticed a progressive shift of the edge of the low-energy band towards higher energies as well as an increasing relative intensity of this band with the covalent character of the chemical bonding through the series FeS $\rightarrow$ CoS $\rightarrow$ NiS. Indeed, the shift of the Fermi level must result in an increase of sulfur occupied states and a corresponding decrease in the unoccupied part of the sulfur p states.

\subsection*{Calculated electronic structure}

\begin{figure}[htb!]
	\centering
    \includegraphics[width=\textwidth]{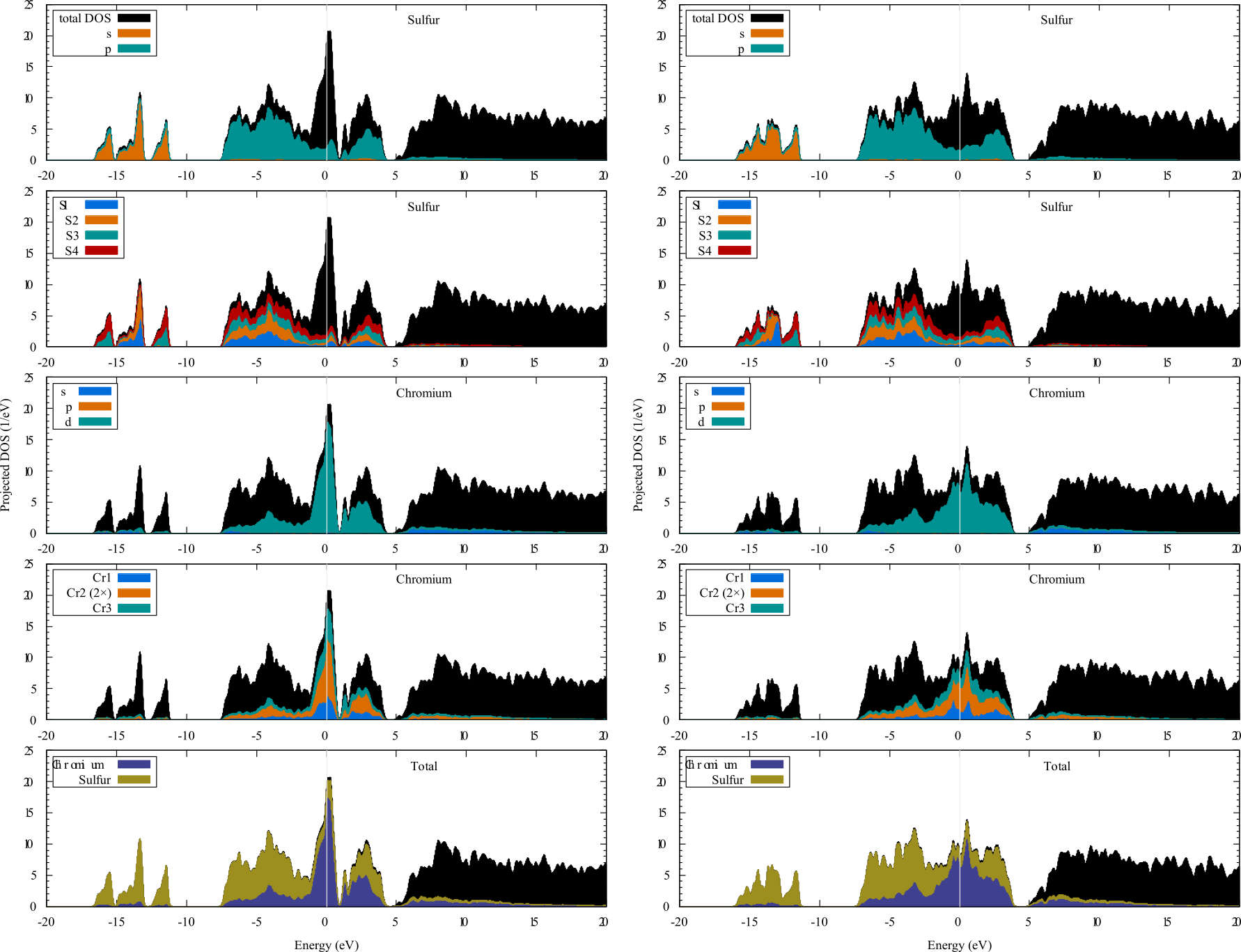}
	\caption{Total and partial densities of states projected onto the different Cr and S sites and onto s, p and d bands for the experimental (left) and relaxed (right) CrS$_2$ structures. The labels of the atomic sites refer to Table \ref{tab:structure}.}
	\centering
	\label{fig:DOS}
\end{figure}

Hereafter, we show that the results of \textit{ab initio} calculations support the previous scenario and provide further indications regarding the electronic structure of the CrS$_2$ phase. In Figure \ref{fig:DOS}, we show the total and partial density of states (DOS) projected onto the sulfur and chromium orbitals for both experimental and relaxed structures. In the experimental structure, note a pronounced metallic behavior, which is reflected by a pronounced peak at the Fermi level, $D(E_F)=$9.07 states eV$^{-1}$cell$^{-1}$ (or 2.27 states eV$^{-1}$ f.u.$^{-1}$). This peak is present in the relaxed structure as well, although less evident. The Fermi energy is located at the center of a large group of bands with covalent character produced by a strong hybridization of 3d Cr and 3p S orbitals.

The projected DOS of the Cr2 and Cr3 sites of Cr (see Figure \ref{fig:structure}) do not differ significantly. This is explained by the fact that both sites are located in the 1T-like 'steps' of the ladder structure of Figure \ref{fig:structure} and therefore probe a similar electronic environment. Indeed, while the pristine (undistorted) 1T structure contains only one octahedrally coordinated transition metal site, in the CrS$_2$ structure, this site is split into two crystallographically inequivalent sites by the monoclinic distortion, still these sites retain a similar structural and chemical environment. The projected DOS of the Cr1 site located in the chain portion of the ladder structure connecting the steps is significantly smaller, which reflects a different structural environment characteristic of the marcasite-type structure. Finally, the four S sites do not display significant differences, which indicates a rather homogeneous charge distribution within the cell, consistent with the metallic properties of the system. 

\subsection*{Transport properties}
The room temperature measurements of the two-terminal resistance on the ten segments of CrS$_2$ wire prepared by FIB as described above yielded values that vary from sample to sample in the 4-300 k$\Omega$ range. We believe that this variability depends not only on the rather different dimensions of the segments but also on the differences in the resistance of the W and contacts and of the contact resistance. As a result, it would be difficult to reliably estimate the resistivity, $\varrho$, of an individual CrS$_2$ nanorod from these resistance measurements at room-temperature. 

\begin{figure}[htb!]
	\centering
    \includegraphics[width=0.8\textwidth]{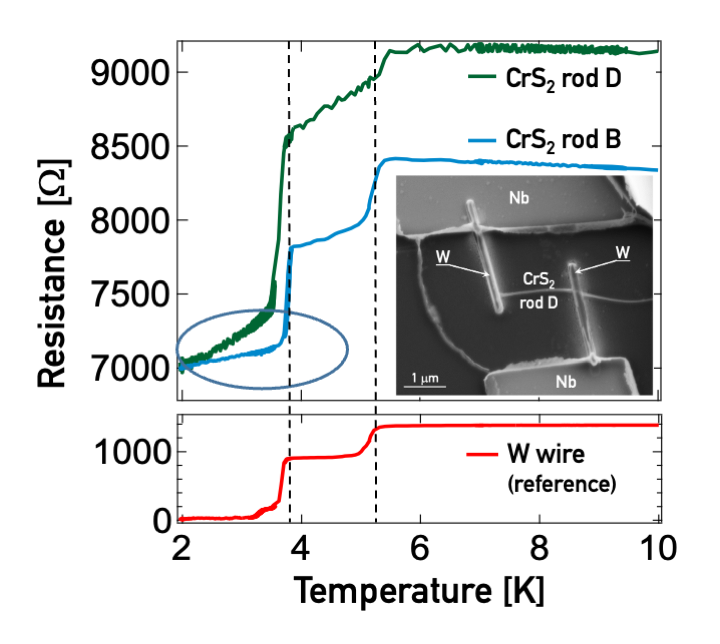}
	\caption{Temperature dependence of the two-terminal resistance of CrS$_2$ rods B and D and of the dummy W wire used as reference sample, as described in the text. The SEM image shows the rod D connected in series with the two bimetallic W/Nb terminals. The vertical broken lines mark the double superconducting transition of these terminals at $T_{c,{\rm Nb}}$=5.5 K and $T_{c,{\rm W}}$=3.7 K. The zero resistance of the reference W wire below $T_{c,{\rm W}}$ indicates that the portion of the curves of rods B and D below this temperature, highlighted by the circle, reflects the intrinsic behavior of CrS$_2$. The filament on the left of the SEM image, a residue of the Nb sputter-deposition through a resist mask, is probably orders of magnitude more resistive than the CrS$_2$ rod, as indicated by the pronounced resistance drop at the superconducting transition of the W electrodes.}
	\centering
	\label{fig:transport}
\end{figure}

In order to overcome this difficulty, we focus on the low-temperature measurements carried out down to 2 K. At these temperatures, both Nb and W are superconducting, so their contribution to the total sample resistance is expected to vanish, thus enabling us to measure the resistance curve, $R(T)$, of the segments of the CrS$_2$ nanorods alone. To do so, as reference, we first measured separately the $R(T)$ curve of a segment of dummy W rod deposited by FIB and connected to a pair of Nb and W contacts, in the same way as for the measurement of the segments of CrS$_2$ wires. The $R(T)$ curve of the dummy W rod is shown in Figure \ref{fig:transport}. As expected, we observe a characteristic step-like behavior with two sharp superconducting transitions at $T_{c,{\rm Nb}} \approx$ 5.5 K and $T_{c,{\rm W}} \approx$ 3.5 K, attributed to Nb and W, respectively, and a vanishing total resistance below $T_{c,{\rm W}}$, where both W and Nb are superconducting. Interestingly, besides the two sharp drops at $T_{c,{\rm Nb}}$ and $T_{c,{\rm W}}$, the curve is flat. Thus, any temperature dependence of the $R(T)$ curves of the CrS$_2$ samples will be safely attributed to an intrinsic transport property of CrS$_2$.

For the low-temperature measurements, we selected two segments of CrS$_2$ rods, labeled B (D), of dimensions 200 (60) nm in diameter and 1.3 (1.7) $\mu$m in length. The $R(T)$ curves of both samples are shown in Figure \ref{fig:transport}. Note a marked positive slope, $dR/dT > 0$, in the two regions between $T_{c,{\rm Nb}}$ and $T_{c,{\rm W}}$ and below $T_{c,{\rm W}}$. This indicates a pronounced metallic behavior of the CrS$_2$ rods, considering that the contribution of the W/Nb terminals to the total resistance is either temperature-independent in the normal state or vanishing in the superconducting state of Nb and W. Assuming the contact resistance between the W terminals and the CrS$_2$ rods to be negligible and by taking into account the rod dimensions given above, we estimate the resistivity of the B and D rods to be $\varrho_{\rm 4 K}$=23 and 2 m$\Omega$ cm, respectively, at 4 K. In the opposite case where the residual resistance of 7 k$\Omega$ at 2 K was mainly due to the contact resistance, the previous estimate should be reduced by one order of magnitude. We attribute the difference between samples B and D to the uncertainty in the estimate of segment dimension and to a difference in the contact resistance. In any case, the order of magnitude of the above estimate of $\varrho$ matches that of other metallic dichalcogenides with positive resistivity coefficients, such as TiS$_2$ and ZrS$_2$ \cite{onu82}.

\section*{Conclusions}
By means of high-pressure and high temperature synthesis, we have synthesized a CrS$_2$ phase with a peculiar ladder-type structure that crystallizes in the form of nanorods. The crystal structure determined experimentally and supported by \textit{ab initio} calculations is formed by two sublattices characteristic of 2D and 3D dichalcogenides with layered 1T- and marcasite-type structures, respectively. Thus, the reported structure bridges the above two families of 2D and 3D structures stable at low and large fillings of the d-band of the transition metal, respectively, and explains the absence of a stable CrS$_2$ at ambient conditions. Our analysis of structural and spectroscopic data indicates that high-pressure synthesis stabilizes an average valence state of Cr$^{4+}$, an unusually high value for chromium sulfides. Electrical resistance measurements give evidence of pronounced metallic properties, consistent with the prediction of \textit{ab initio} calculations. Further studies would be desirable to verify the possibility of obtaining macroscopic samples of the present CrS$_2$ phase. Such samples would enable to investigate in detail the electronic and transport properties and test the potential of CrS$_2$ as functional material for energy-storage or catalysis applications that may be possible owing to the presence of channels in the crystal structure.

\begin{acknowledgement}

The authors thank Fran\c cois Guyot for useful discussions and Im\`ene Est\`eve for assistance in the SEM characterization of the samples. They gratefully acknowledge the French platform METSA for the assigned microscope time at the IRMA CRISMAT laboratory to perform the PEDT and atomic STEM imaging experiments (Project METSA 14B111).

\end{acknowledgement}

%\begin{suppinfo}

%This will usually read something like: ``Experimental procedures and
%characterization data for all new compounds. The class will
%automatically add a sentence pointing to the information on-line:

%\end{suppinfo}

\bibliography{CrS2}

\end{document}